\documentclass{article}
\usepackage{spconf,amsmath,graphicx}
\usepackage[table,xcdraw]{xcolor}

\title{A Novel Monocular Disparity Estimation Network with Domain Transformation and Ambiguity Learning}
\name{Juan Luis Gonzalez Bello and Munchurl Kim}
\address{Korea Advanced Institute of Science and Technology (KAIST)}
 
\begin{document}
\ninept
\maketitle
\begin{abstract}
Convolutional neural networks (CNN) have shown state-of-the-art results for low-level computer vision problems such as stereo and monocular disparity estimations, but still, have much room to further improve their performance in terms of accuracy, numbers of parameters, etc. Recent works have uncovered the advantages of using an unsupervised scheme to train CNN's to estimate monocular disparity, where only the relatively-easy-to-obtain stereo images are needed for training. We propose a novel encoder-decoder architecture that outperforms previous unsupervised monocular depth estimation networks by (i) taking into account ambiguities, (ii) efficient fusion between encoder and decoder features with rectangular convolutions and (iii) domain transformations between encoder and decoder. Our architecture outperforms the Monodepth baseline in all metrics, even with a considerable reduction of parameters. Furthermore, our architecture is capable of estimating a full disparity map in a single forward pass, whereas the baseline needs two passes. We perform extensive experiments to verify the effectiveness of our method on the KITTI dataset. 
\end{abstract}


\section{Introduction}
\label{sec:intro}
For a given object displayed in a rectified stereo pair of images, the disparity is defined as the horizontal pixel distance of the object in the left and right images. Disparity estimation is an ill-posed problem due to occlusion, as not all the objects displayed in the left image are visible in the right image and vice versa. Classic approaches for depth / disparity estimation rely on stereo matching, multiple view stereo, or single or multiple defocus map \cite{Malik2008ANA}. These methods rely on multiple images to perform feature matching, triangulation, or texture measurement with limited performance. On the other hand, even early supervised learning approaches have demonstrated superior results \cite{ZbontarL15, neuralfields} for binocular and monocular inputs respectively.

\subsection{Supervised disparity estimation}
Deep learning approaches have been extensively studied for supervised disparity estimation. For the stereo inputs case, the early work of Zbontar \cite{ZbontarL15} focused on learning the similarity measure between two patches from the left and right images. For the monocular case, Liu et al. \cite{neuralfields} devised a deep convolutional neural field (DCNF) model for supervised depth estimations by exploring conditional random fields and super-pixel pooling. None of these previous methods provided an end-to-end learning architecture. Instead of comparing local patches, Mayer et al.\cite{mayer} adopted the fully-convolutional FlowNet \cite{flownet} architecture for dense supervised stereo matching, called the DispNet. This kind of auto-encoder architecture would become common for future disparity estimation networks. Jie et al. \cite{recsm} proposed a recurrent architecture that learns potentially erroneous areas that guide the model to focus on these regions for subsequent refinement. Their model is fully supervised, takes stereo inputs and needs 5 iterations during test time, making it too slow for real-time (4 seconds). Atapour et al. \cite{adastyle} leverage fully annotated synthetic data to train a monocular depth estimation network. They apply style transfer to convert natural images into the synthetic domain during test time. While their approach unlocks the use of big synthetic data for real applications, according to their paper, their method struggles from false objects and depth holes arising from shadows post style transfer. Xie et al. \cite{singleviewsm} combine monocular and stereo input approaches using a Deep3D model to generate a synthetic right view, followed by 1D correlation, and a full resolution DispNet architecture to perform stereo matching. Deep3D \cite{deep3d} produces a probabilistic disparity map to blend multiple versions of the left view with different left and right shifts.

Recent works on disparity estimation have the tendency to combine other low-level and high-level tasks like motion estimation, scene flow, and object segmentation \cite{occmondep, segstereo}. Chen et al. \cite{stereo_style} formulate the style transfer problem for the binocular input case. Their intermediate disparity is trained for simultaneous bidirectional disparity and occlusion mask estimation in a fully supervised fashion.

\begin{figure}
  \centerline {\begin{tabular}{p{2.4cm}p{2.4cm}p{2.4cm}}
  Training inputs & Disparity maps & Ambiguity masks
  \end{tabular}}
  \includegraphics[width=0.48\textwidth]{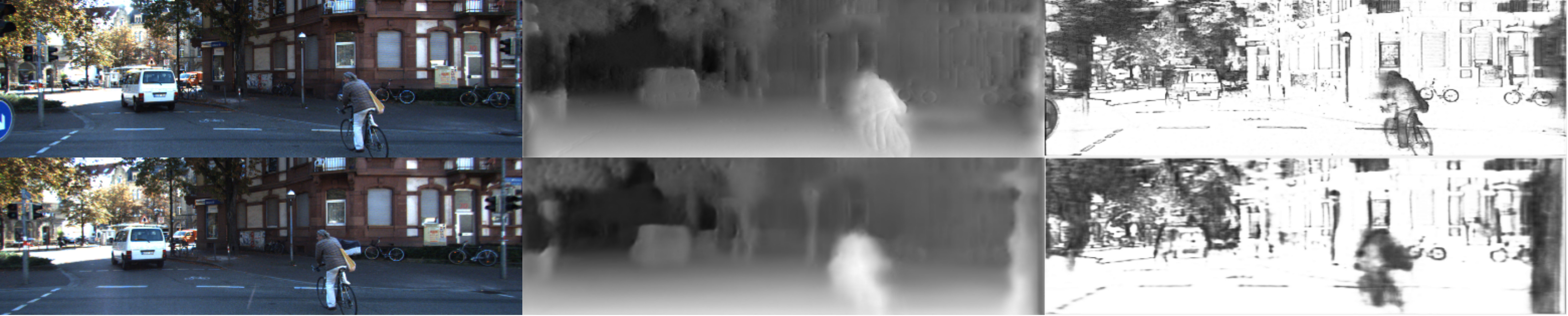}
  \vspace*{-7mm}
  \caption{Using the left view as input and the right view only for supervision during training, our model estimates bidirectional disparity and ambiguity masks.}
  \label{fig:inputoutput}
\end{figure}

\begin{figure*}
  \centering
  \includegraphics[width=\textwidth]{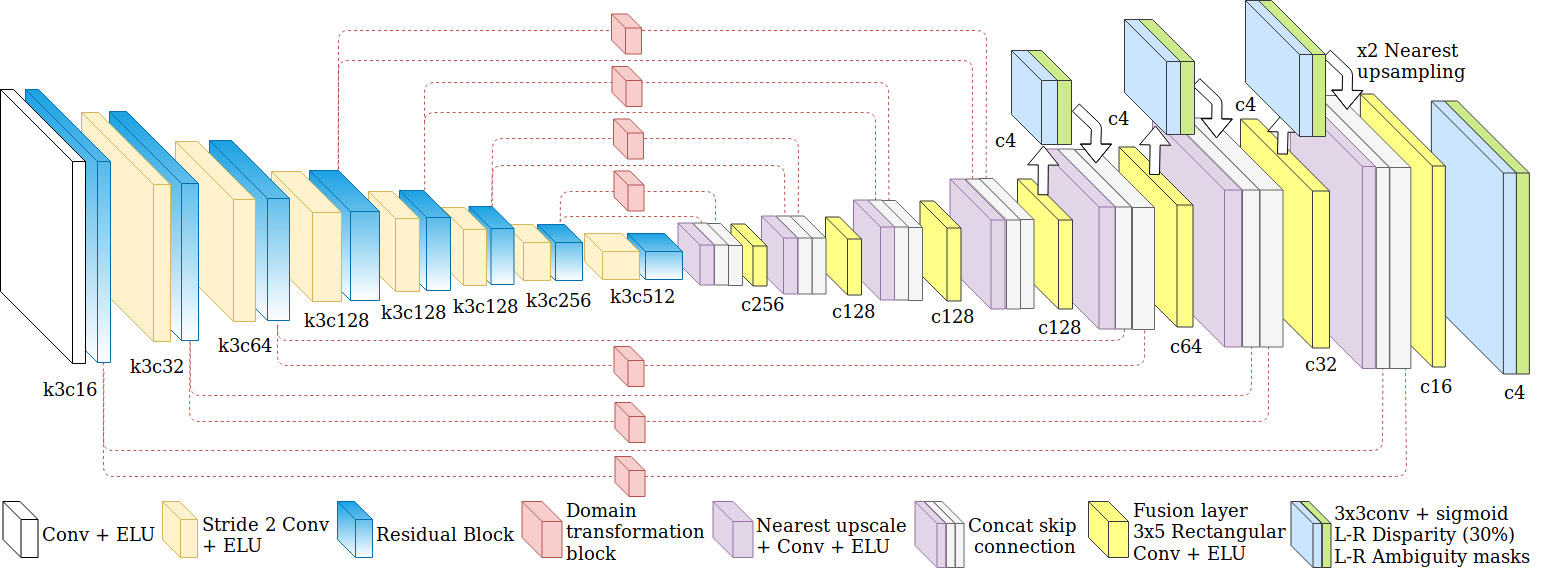}
  \vspace*{-8mm}
  \caption{Our full model with ambiguity mask estimation, rectangular convolutions and domain transformation blocks.}
  \label{fig:model}
\end{figure*}

\subsection{Unsupervised disparity estimation}
\label{ssec:subhead}
Previously mentioned methods require dense ground truth disparity. Obtaining such annotated data is a challenging task. On the other hand, unsupervised techniques rely on additional views to estimate depth from a scene, for which, capturing monocular video or stereo images is a relatively simpler task. Zhou et al.\cite{sfmlearner} exploited the relative pose information in a monocular video to train a disparity network and a pose estimation network in an unsupervised manner by performing view synthesis of the center frame (source image) to the different reference frames (target images). In \cite{unflow}, Zou et al. proposed a cross-task consistency loss to jointly train for optical flow estimation and disparity estimation from a video in an unsupervised fashion. Even though these kinds of approaches achieve good results for the additional task (camera motion and optical flow estimations), their performance for disparity estimation is very limited.

Given only the stereo pair, unsupervised disparity estimation can benefit from exploiting the geometrical relationships between the left and right images. Garg et al. \cite{garg} trained an encoder-decoder architecture for unsupervised monocular depth estimation by synthesizing a backward-warped image using the estimated left disparity map and the right image. The warped images are used to calculate the reconstruction error. We use the state-of-the-art work of Godard et al. \cite{godard} as the baseline for our work. Their Monodepth network estimates the left and right view disparity maps and incorporates a photometric loss, a disparity smoothness loss and a consistency loss for unsupervised training. Even though their consistency loss term greatly improves the performance of the network, they fail in the following three points that we attack in our work, which we call the rrdispnet\_dtm (residual rectangular masked disparity network with left to right domain transformations):

(i) Incorporation of full resolution features from the encoder into the decoder. We add a convolution layer with full resolution features to the auto-encoder architecture;
(ii) Fusion between “Left-domain” encoder features into “Left-Right-domain” decoder features. We fuse the skip connections from the encoder into the decoder using domain transformation blocks and rectangular convolutions with 3x5 kernels instead of 3x3 kernels. Rectangular convolutions facilitate the conversion from Left-domain to Left-Right-domain;
(iii) Accounting for "ambiguities" in the loss function. By letting the network learn an ambiguity mask for each view, we effectively re-weight the loss functions, which results in a selectively decreased learning rate for occluded and complex cluttered disparity areas, which improves accuracy and robustness. In addition, accounting for ambiguities allows our network to estimate full disparity maps in a single pass.

These contributions greatly improve the quality of the disparity maps. Moreover, in combination with residual blocks \cite{resnet} in the encoder section, we can reduce the numbers of parameters from 31 to 14 million. Parameter reduction is achieved by having less numbers of channels in the intermediate stages of our auto-encoder architecture, we trade quantity for quality of features. 

\section{Method}
\label{sec:method}
We propose a novel learning pipeline that accounts for occlusions and complex/cluttered areas or "ambiguities". Given a single image, our model outputs bidirectional disparity and ambiguity masks as depicted in Figure \ref{fig:inputoutput}. The works of Godard \cite{godard} and Garg \cite{garg} model depth estimation as image reconstruction by $\Tilde{I}_L = g(I_R, D_L)$, where the backward warping operation $g(I,D)$ is a fully differentiable bilinear sampling function for right image $I_R$ and left disparity $D_L$. However, these approaches do not take ambiguities into account. Our  pipeline handles ambiguities by including them in the reconstruction model, as depicted in Eqs.\ref{eq:lrec} and \ref{eq:rrec} for left and right view reconstruction respectively.
\begin{equation} \label{eq:lrec}
\Tilde{I}_L = \Tilde{a}\_mask_L \odot g(I_R, D_L) + (1 - \Tilde{a}\_mask_L) \odot I_L
\end{equation}
\begin{equation} \label{eq:rrec}
\Tilde{I}_R = \Tilde{a}\_mask_R \odot g(I_L, D_R) + (1 - \Tilde{a}\_mask_R) \odot I_R
\end{equation}
where $\Tilde{a}\_mask_L$ and $\Tilde{a}\_mask_R$ contain the information about the dis-occluded left and right border areas that cannot be reconstructed by the warping operation. To model these areas the ambiguity masks are element-wise multiplied, denoted as $\odot$, by the dis-occlusion masks, yielding $\Tilde{a}\_mask_L = a\_mask_L \odot dis\_occ_L$ and $\Tilde{a}\_mask_R = a\_mask_R \odot dis\_occ_R$. The later are defined as
\begin{equation} \label{eq:disocc_maskL}
dis\_occ_{Lij} =
\left \{
  \begin{tabular}{cc}
  $0$ & $if\ j < 0.15W$ \\
  $1$ & $o.w.$ \\
  \end{tabular}
\right.
\end{equation}
\begin{equation} \label{eq:disocc_maskR}
dis\_occ_{Rij} =
\left \{
  \begin{tabular}{cc}
  $0$ & $if\ j > 0.85W$ \\
  $1$ & $o.w.$ \\
  \end{tabular}
\right.
\end{equation}
where $W$ is the image width.

\begin{figure}
  \centering
  \includegraphics[width=0.35\textwidth]{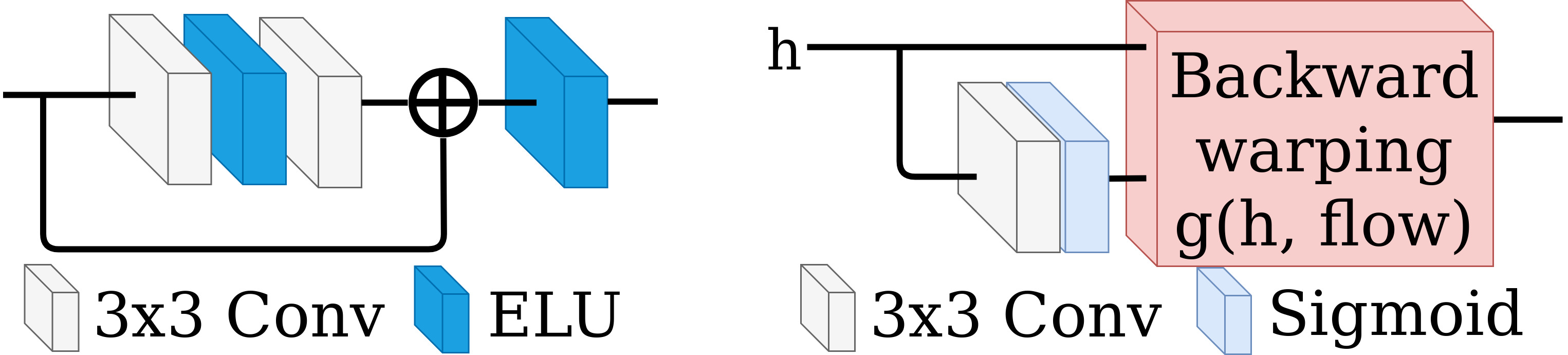}
  \vspace*{-3mm}
  \caption{Residual(left) and domain transform(right) blocks.}
  \label{fig:blocks}
\end{figure}

\subsection{Network architecture}
\label{ssec:loss}
Our network architecture is depicted in Figure \ref{fig:model} where we adopt a UNET-like architecture with several additions. After each strided convolution layer, a residual block (Figure \ref{fig:blocks}) further refines the features and increases the numbers of non-linearities in the network, increasing its representation power. The bottleneck features in our network are upscaled using nearest upsampling and concatenated with encoder features in order to include global and local information. The features from the encoder are concatenated to the decoder side via direct skip connections and domain transformations. The number of channels is then reduced by a fusion layer. This process repeats until the feature maps reach the input resolution. Since the encoder processes spatial information from the left input image and the decoder outputs left and right disparities and masks, a mechanism is needed to convert from the "Left domain" encoder features to the "Left-Right" domain decoder features. The domain transformation blocks (Figure \ref{fig:blocks}) and 3x5conv fusion layers facilitate the conversion from the Left domain features to the Left-Right domain features. Similar to \cite{godard}, we take multiscale outputs for training at the four finest scales. The intermediate outputs are upscaled using nearest upsampling and concatenated to the next decoder stage. Output disparities and ambiguity masks are preceded by a sigmoid activation. Disparities are limited to $30\%$ of the image width.

\subsection{Loss Functions}
\label{ssec:loss}
We adopt a multi-scale loss function at the four finest scales as described by Eq. \ref{eq:mscale}. For each scale, the loss function takes the shape of the weighted sum of five terms: reconstruction loss, edge preserving smoothness loss, perceptual loss, ambiguity loss, and left-right consistency loss. The loss function at each scale is depicted in Eq. \ref{eq:one_scale} (each term has a right and left component $l_x^L$ and $l_x^R$).
\begin{equation} \label{eq:mscale}
l = \sum_{s=0}^{4} l_s
\end{equation}
\begin{equation} \label{eq:one_scale}
l_s = a_{rec}l_{rec} + a_{ds}\dfrac{0.1}{2^{s-1}}l_{ds} + a_pl_p + a_al_a + a_{lr}l_{lr} 
\end{equation}
The selection of the coefficients for ambiguity and perceptual loss terms in Eq. \ref{eq:one_scale} was critical for the correct training of the networks. The weights during training are set to $a_{rec}=1$, $a_{ds}=0.1$, $a_p=0.1$, $a_a=0.2$ and $a_{lr}=1$. In contrast with \cite{godard}, all our loss terms are directly or indirectly modified by the ambiguity masks.
\textbf{Reconstruction Loss}. The reconstruction loss enforces the image $\Tilde{I}_L$ to be similar to the input image $I_L$, and can be defined by the weighted sum of the $L1$ and $SSIM$ losses (a weight of $\alpha = 0.85$ was used). The reconstruction loss is defined as
\begin{equation} \label{eq:rec_loss}
l_{rec}^L = \alpha||I_l - \Tilde{I}_L||_1 + (1 - \alpha)SSIM(I_L, \Tilde{I_L})
\end{equation}
\textbf{Disparity smoothness Loss}. Similar to \cite{godard}, the smoothness loss is set up to put less penalty on image gradients, which is given by 
\begin{equation} \label{eq:smooth_loss}
l_{ds}^L = ||\partial_x D_L \odot \exp^{-|\partial_x I_L|}||_1 + ||\partial_y D_L \odot \exp^{-|\partial_y I_L|}||_1
\end{equation}
\textbf{Perceptual Loss}. Occluded areas will normally be represented by highly deformed regions in the reconstructed images. Perceptual loss \cite{perceptual} is ideal to detect this deformation and to put more penalty on it. The use of perceptual loss allows for learning the ambiguity masks properly. Three layers ($relu1\_2, relu2\_2, relu3\_4$) from the pre-trained $VGG19$ \cite{vgg} on ImageNet were used to have our perceptual loss as
\begin{equation} \label{eq:vgg_loss}
l_p^L = \sum_{l=1}^{3} ||\phi^l(I_L)-\phi^l(\Tilde{I_L})||_1
\end{equation}
\textbf{Ambiguity Loss}. Cross entropy loss is used to encourage ambiguity mask elements to be close to 1. Without this term the ambiguity masks would collapse to zero. So we have the ambiguity loss as
\begin{equation} \label{eq:a_loss}
l_a = ||\log a\_mask_L||_1 + ||\log a\_mask_R||_1
\end{equation}
\textbf{Left-Right (LR) consistency Loss}. Similar to \cite{godard}, a LR consistency term is used. Consistency loss encourages coherence between left and right disparities. Another interpretation is that consistency loss allows the network to use each view's disparity map as weak ground truth for the other view. In contrast with \cite{godard}, our LR consistency loss is re-weighted by the $\Tilde{a}\_mask$ ambiguity masks. The ambiguity masks allow the LR consistency loss to penalize less those areas where disparities are not good enough to be used as the ground truth for the other view's disparity. The LR consistency loss is defined as
\begin{equation} \label{eq:rec_loss}
l_{lr}^L = ||\Tilde{a}\_mask_L \odot (D_L - g(D_R, D_L))||_1
\end{equation}

\begin{table*}[t]
    \centering
    \begin{tabular}{|c|l|c|c|c|c|c|c|c|c|c|c|c|c|}
\hline
\rowcolor[HTML]{C0C0C0} 
Model                                  & D                      & R                        & F                        & abs rel                      & sq rel                       & rmse                         & log rmse                     & a1                           & a2                           & a3                           & Warp rmse                     & Time                         & Param                       \\ \hline
Monodepth                              &                        &                          &                          & 0.149                        & 2.565                        & 6.645                        & 0.245                        & 0.849                        & 0.936                        & 0.969                        & 17.565                        & 0.015                        & 31.6                        \\ \hline
Monodepth pp                           &                        &                          & x                        & 0.114                        & 1.138                        & 5.452                        & 0.204                        & 0.859                        & 0.946                        & 0.977                        & 17.565                        & 0.032                        & 31.6                        \\ \hline
rdispnet\_m                            &                        &                          & x                        & \textbf{0.111}               & \textbf{1.031}               & 5.416                        & 0.199                        & 0.860                        & 0.948                        & 0.978                        & 17.244                        & \textbf{0.014}               & \textbf{12.8}               \\ \hline
rrdispnet\_m                           &                        & x                        & x                        & 0.113                        & 1.114                        & 5.364                        & \textbf{0.195}               & \textbf{0.866}               & \textbf{0.951}               & \textbf{0.981}               & 17.062                        & 0.018                        & 14.2                        \\ \hline
\textbf{rrdispnet\_dtm}                & \multicolumn{1}{c|}{x} & x                        & x                        & 0.112                        & 1.038                        & \textbf{5.304}               & 0.198                        & 0.863                        & 0.950                        & 0.979                        & \textbf{16.791}               & 0.024                        & 16.0                        \\ \hline
\rowcolor[HTML]{EFEFEF} 
{\color[HTML]{000000} rrdispnet\_m pp} &                        & {\color[HTML]{000000} x} & {\color[HTML]{000000} x} & {\color[HTML]{000000} 0.105} & {\color[HTML]{000000} 0.949} & {\color[HTML]{000000} 5.174} & {\color[HTML]{000000} 0.190} & {\color[HTML]{000000} 0.866} & {\color[HTML]{000000} 0.952} & {\color[HTML]{000000} 0.981} & {\color[HTML]{000000} 17.062} & {\color[HTML]{000000} 0.036} & {\color[HTML]{000000} 14.2} \\ \hline
\end{tabular}
    \vspace*{-3mm}
    \caption{Network Performance Metrics. D, network uses domain transformation block; R, network uses rectangular convolutions; F, network estimates full disparity map. Inference time is in seconds (s), tested on a Titan Xp GPU. Numbers of parameters is in millions.}
    \label{tab:performance}
\end{table*}

\begin{figure}
  \centering
  \includegraphics[width=0.48\textwidth]{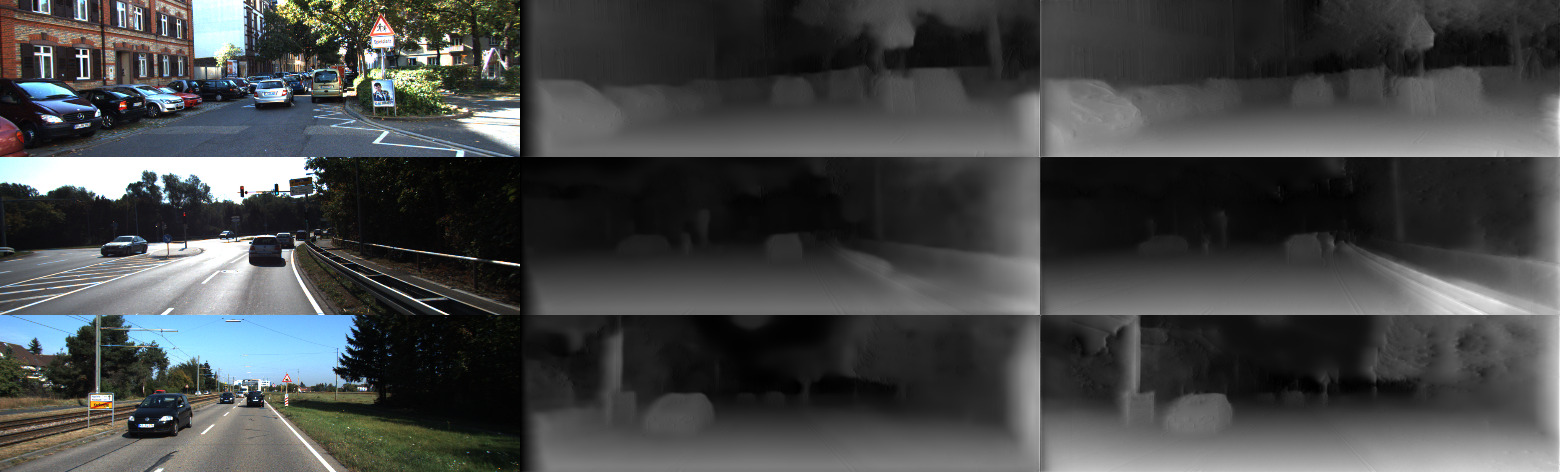}
  \vspace*{-8mm}
  \caption{From left to right: Input, Incomplete disparity map (Monodepth) and our complete result (rdispnet\_m).}
  \label{fig:monovsours}
\end{figure}

\section{Results}
\label{sec:results}
We perform extensive experiments to verify the effectiveness of each of our contributions and compare against the state-of-the-art unsupervised monocular depth estimation, the Monodepth \cite{godard}, on the Kitti2015 \cite{kitti2015} dataset which contains 200 stereo frames and sparse disparity ground truth from velodyne laser scanners and CAD models. We train and test our network with and without domain transformations and rectangular convolutions, and name them accordingly, as shown in Table \ref{tab:performance}. Performance is measured in terms of the Kitti metrics from \cite{eigen}. Additionally, we test the Monodepth pp and our best model on our own dataset, captured with a cellphone camera.

\begin{figure*}
  \centering
  \includegraphics[width=0.96\textwidth]{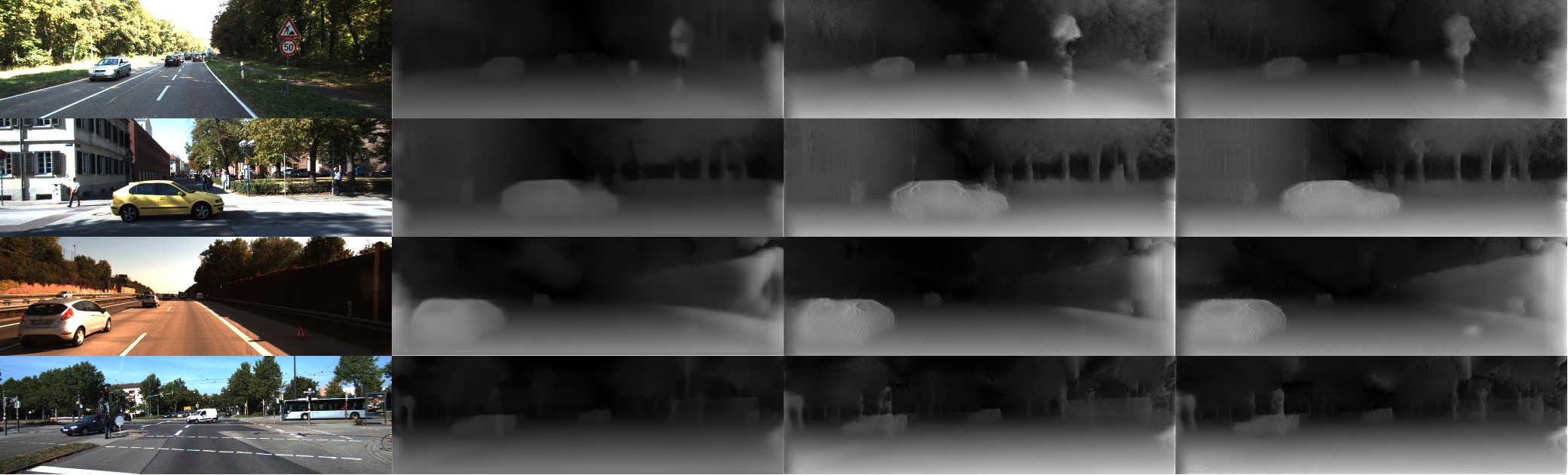}
  \vspace*{-4mm}
  \caption{Qualitative comparison on the Kitti2015 dataset. From left to right: Input, Monodepth pp, rrdispnet\_m and rrdispnet\_dtm. Our networks succeed at detecting thin structures and suffer less from lateral side artifacts.}
  \label{fig:qualitative}
\end{figure*}

\subsection{Implementation details}
For a fair comparison, we adopted the training settings from \cite{godard}. Adam \cite{adam} optimizer is used with default betas and the models are trained for 50 epochs with an initial learning rate of 0.0001. The learning rate is halved at epochs 30 and 40. The same data augmentations were performed: random crop (256x512), random horizontal flips, random gamma, brightness and color shifts \cite{godard}. All models are trained on the KITTI Split \cite{godard} only, which consists of a selection of 29,000 stereo pairs from the Kitti2012 dataset \cite{kitti2012}, comprising a total of 33 scenes.

\subsection{Kitti2015}
\begin{figure}
  \centering
  \includegraphics[width=0.38\textwidth]{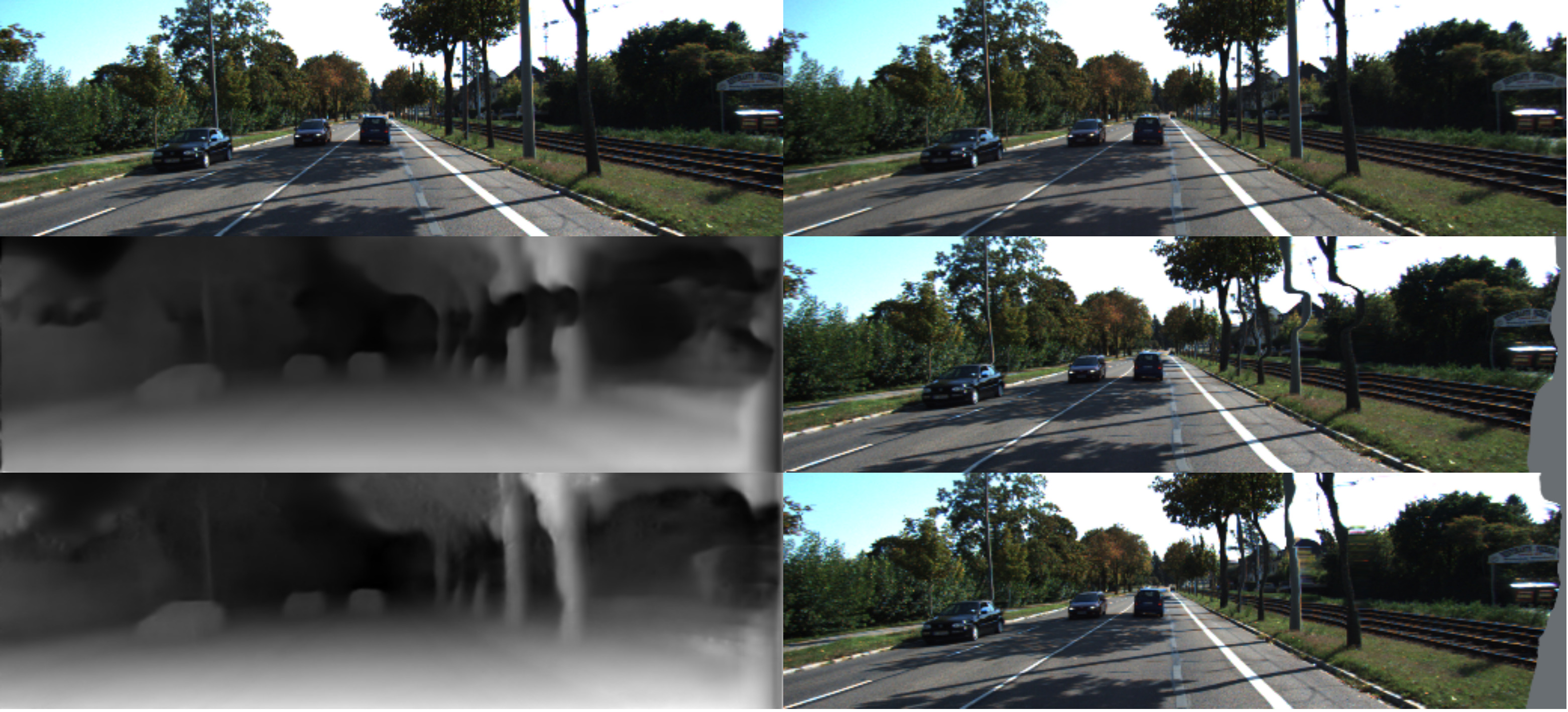}
  \vspace*{-4mm}
  \caption{Left column: Input left view, Monodepth and rrdispnet\_dtm output right disparities. Right column: Ground truth right view, warped left view using disparities in left column.}
  \label{fig:rightview}
\end{figure}

Our simplest network, rdispnet\_m, does not include domain transformations nor 3x5 convolutions, but still manages to outperform the Monodepth baselines in all metrics. Monodepth post processing (pp), runs a second forward pass with a flipped input and combines the two outputs into a full disparity map. In contrast, our network is able to generate a complete disparity map in a single forward pass as depicted in Figure \ref{fig:monovsours}, even with the higher overall quality, as confirmed by results in Table \ref{tab:performance}. 
We compare our more complex networks with rectangular convolutions and domain transformations against Monodepth pp in Figure \ref{fig:qualitative}. Again, our networks outperform Monodepth pp in all metrics in Table \ref{tab:performance}. The additional complexity of 3x5 kernels and domain transformations makes our rrdispnet\_dtm slower than our rdispnet\_m but still faster than Monodepth pp. The rrdispnet\_dtm yields the lowest rmse and has intermediate values for the other Kitti metrics, which exhibits the best performance among all our networks. As can be observed in Figure \ref{fig:qualitative}, our rrdispnet\_dtm generates sharper and more accurate disparity maps. It is also more robust against thin structures and does not produce lateral side artifacts as the Monodepth pp does. In addition, the rrdispnet\_dtm has the lowest warp rmse, which tells how similar the generated right-from-left view to the ground truth right view is. The warp rmse is a rough indicator for the quality on the estimated right disparity. The estimated right disparities and the generated right views are depicted in Figure \ref{fig:rightview}. Interestingly, our network does a good job at estimating the full right disparity map under the case that the right view is unknown at test time.
\subsection{Own dataset}
\begin{figure}
  \centering
  \includegraphics[width=0.47\textwidth]{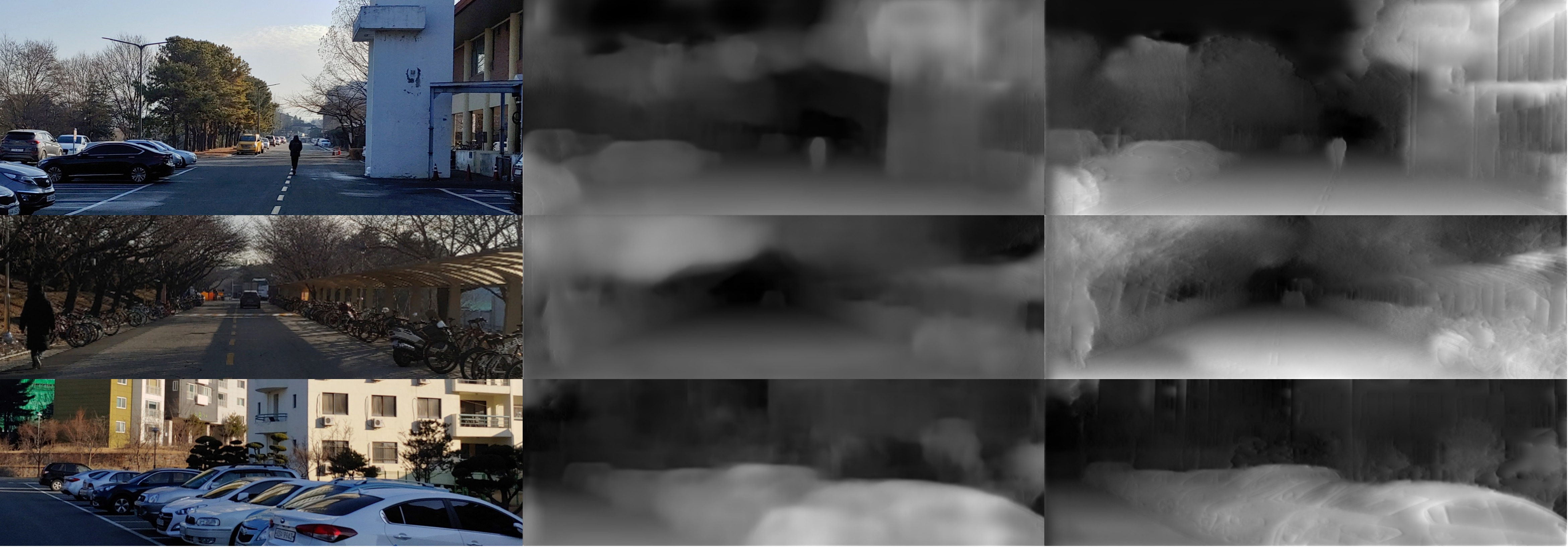}
  \vspace*{-4mm}
  \caption{From left to right: Input, Monodepth pp and rrdispnet\_dtm.}
  \label{fig:own}
\end{figure}

To prove our models generalize well, we test the rrdispnet\_dtm on our own dataset, and compare it against the Monodepth pp baseline (Figure \ref{fig:own}). Since the ground truth disparity is not available for our dataset, we evaluate the quality of the disparity maps by the feasibility of the generated outputs. Our network generates more feasible disparity maps where details are very well preserved, thin structures better detected, and far away object disparities better estimated.

\section{Conclusions}
\label{sec:conclusions}
In this paper, by using residual blocks, full resolution features, rectangular convolution fusion and domain transformations between the Left-domain and Left-Right-domain features combined with the estimations of ambiguity masks, we achieved superior qualitative and quantitative results on the Kitti2015 benchmark over the Monodepth baseline. Furthermore, using our own dataset we showed that our method generalizes better in comparison to the baseline. The design of our novel loss function allows for end-to-end unsupervised learning of binocular disparity and ambiguity masks. We presented three networks that generate full disparity maps in a single pass at higher speeds than the Monodepth pp baseline. While other recent works are focused on fully supervised disparity networks, we demonstrated a significant improvement in the unsupervised class of algorithms by modeling depth estimation as an ambiguous image reconstruction problem. 

\bibliographystyle{IEEEbib}
\bibliography{refs}

\end{document}